\documentclass[aps,epsfig,showpacs,twocolumn,letterpaper]{revtex4}
\usepackage{graphicx}
\usepackage{amsmath}
\usepackage{latexsym}
\usepackage{amsfonts}
\usepackage{amssymb}
\usepackage{array}
\usepackage{color}
\usepackage{subfigure}
\usepackage{verbatim}

\begin{document}
\newcommand{\Q}[1]{{\color{red}#1}}
\newcommand{\blue}[1]{{\color{blue}#1}}
\newcommand{\red}[1]{{\color{red}#1}}
\newcommand{\Change}[1]{{\color{green}#1}}
\newcommand{\av}[1]{\langle#1\rangle}
\newcommand{\Matrix}[1]{\left(\begin{array}{ccc}#1\end{array}\right)}

\title{Direct observation of phase sensitive Hong-Ou-Mandel interference}
\author{Petr Marek}
\affiliation{
Department of Optics, Palack\'y University, 17. listopadu 1192/12, 77146 Olomouc, Czech Republic}
\author{Petr Zapletal}
\affiliation{
Department of Optics, Palack\'y University, 17. listopadu 1192/12, 77146 Olomouc, Czech Republic}
\author{Radim Filip}
\affiliation{
Department of Optics, Palack\'y University, 17. listopadu 1192/12, 77146 Olomouc, Czech Republic}
\author{Yosuke Hashimoto}
\affiliation{Department of Applied Physics and Quantum-Phase Electronics Center, School of Engineering, The University of Tokyo, 7-3-1 Hongo, Bunkyo-ku, Tokyo 113-8656, Japan}
\author{Takeshi Toyama}
\affiliation{Department of Applied Physics and Quantum-Phase Electronics Center, School of Engineering, The University of Tokyo, 7-3-1 Hongo, Bunkyo-ku, Tokyo 113-8656, Japan}
\author{Jun-ichi Yoshikawa}
\affiliation{Department of Applied Physics and Quantum-Phase Electronics Center, School of Engineering, The University of Tokyo, 7-3-1 Hongo, Bunkyo-ku, Tokyo 113-8656, Japan}
\author{Kenzo Makino}
\affiliation{Department of Applied Physics and Quantum-Phase Electronics Center, School of Engineering, The University of Tokyo, 7-3-1 Hongo, Bunkyo-ku, Tokyo 113-8656, Japan}
\author{Akira Furusawa}
\affiliation{Department of Applied Physics and Quantum-Phase Electronics Center, School of Engineering, The University of Tokyo, 7-3-1 Hongo, Bunkyo-ku, Tokyo 113-8656, Japan}

\begin{abstract}
Quality of individual photons and their ability to interfere is traditionally tested by measuring the Hong-Ou-Mandel photon bunching effect. However, this phase insensitive measurement only tests the particle aspect of the quantum interference.
Motivated by these limitations, we formulate a witness capable of recognizing both the quality and the coherence of two photons from a single set of measurements.
We exploit the conditional nonclassical squeezing which is sensitive to both the quality of the single photons and their indistinguishability and we show that it can be used to directly reveal both the particle and the wave aspects of the quantum interference.
Finally, we experimentally test the witness by applying it to a pair of independent single photons generated on demand.

\end{abstract}


\maketitle
Quantum interference of individual photons is one of the fundamental keystones of quantum technology. The ability to recognize whether photons from different sources can interfere is therefore an important one. The basic tests use photon correlation measurements to ascertain presence of individual photons and two-photon bunching to verify their interference. The experimental test of the bunching effect traditionally employs the Hong-Ou-Mandel interference \cite{mandel}, in which the two individual photons are mixed on a balanced beam splitter and the resulting two mode field is measured by a pair of photon counters. The bunching manifests as the lack of coincidence counts - if the coincidence probability is below one half, the interference effect is nonclassical, i.~e. incompatible with description using a mixture of classical waves or classical particles. Recent experiments for a wide range of experimental platforms safely observed low values of coincidence counts and thus demonstrated this fundamental nonclassical aspect \cite{DVexperiments}.

For discrete variable quantum technology (DV) these features are sufficient. However, from the point of view of continuous variables (CV) and the more general framework of hybrid quantum technology \cite{hybrid,ourjoumtsev,Etesse,hybrid2}, these experiments quantify only the particle nature of the interference and ignore the vacuum and the related phase sensitive aspects \cite{zapletal} of the photon bunched state. The full picture of two-photon interference can be obtained by replacing the photon counters with homodyne detection and performing a full tomography \cite{MakinoHOM,tomography2}. The reconstructed density matrix can then be analyzed in order to visualize and evaluate the interference caused by the the indistinguishability. The drawback to this tomographic approach is that it can not be used as a direct witness.  The tomographic reconstruction is a process combining information from many incompatible measurements in order to obtain not directly measurable quantities, such as off-diagonal elements of the density matrix.

The elements of the reconstructed matrix contain, in principle, full information about the state. However, the information is practically limited due to unavailability of a complete set of measurement and the need to form prior assumptions about the suitable Hilbert space of the system \cite{hradil}. For this reason some elements, in particular those related to higher Fock numbers, are difficult to estimate with sufficient precision. This may lead to erroneous or indeterminate conclusions, such as attributing nonclassical properties to interference of classical signals.
Fortunately, there are some phase sensitive photon correlation properties which can be used for construction of a direct witness not reliant on the tomographic approach.

Two-photon interference leads to a state which, by suitably conditioning homodyne measurement, can be transformed into a state with quadrature squeezing \cite{zapletal}. This squeezing, although not strong enough to be of much use on its own, can be created neither from classical light nor from non-interfering photons and is therefore an excellent witness of nonclassical and coherence properties.
It can be also directly measured.
Interestingly, the measurement itself relies only on classical correlations within the state, without exploiting its entanglement \cite{entanglement} or discord \cite{discord}. The general witness therefore requires one further condition - the phase of the initial states needs to be randomized. This step, while not affecting single photon states, cancels the possible false positives. The full \textit{phase sensitive Hong-Ou-Mandel effect} therefore serves as a directly observable witness simultaneously for both particle and wave-like aspects of the two photon interference. It also offers a new, much more challenging, benchmark against which new single photon sources can be tested.

In this paper we introduce the phase sensitive Hong-Ou-Mandel effect as the witness for both high quality and high overlap of remotely prepared single photon states. The witness relies on observation of nonclassical quadrature correlations in a state created by interference of two single photons. We apply the witness to a pair of photons individually prepared in timing synchronization cavities \cite{MakinoHOM}, and we demonstrate both their high quality and coherence.

In the ideal version of the Hong-Ou-Mandel experiment, the two single photons indistinguishable in all but a single degree of freedom are in the initial state $|1,1\rangle$ and get mixed on a balanced beam splitter. This produces an entangled state
\begin{equation}\label{state_id}
    |\psi\rangle = \frac{1}{\sqrt{2}}(|20\rangle - e^{i\phi}|02\rangle),
\end{equation}
with $\phi$ being the relative phase between the two output modes introduced by the beam splitter.
The two resulting modes are then subjected to intensity measurements by avalanche photo diodes (APDs) whose action is described by their POVM elements $\Pi_0 = |0\rangle\langle 0|$ and $\Pi_1 = 1 - \Pi_0$, which provide us with probabilities $P_{kj} = \langle \psi|\Pi_k\otimes\Pi_j|\psi\rangle$, $k,j = 0,1$. For indistinguishable photons there is a distinct lack of coincidences represented by $P_{11}=0.$ On the other hand, for completely distinguishable photons the probability of coincident detection is firmly set to $P_{11} = 1/2$. The full Hong-Ou-Mandel measurement is realized by deliberately delaying one of the photons, thus increasing the distinguishability, and measuring the coincidence rates for these different settings. The effect is then quantified in terms of visibility, which is defined as
\begin{equation}\label{}
    V = \frac{\max P_{11} - \min P_{11}}{\max P_{11} + \min P_{11}},
\end{equation}
where the maximization and minimization is performed over the set of different delays, and it is equal to one for perfectly indistinguishable single photons and zero for perfectly distinguishable ones. The visibility can be also reduced due to imperfection of the single photon states, caused by presence of vacuum or higher Fock terms. However, since the vacuum terms do not contribute towards coincidence rates, only the higher Fock terms lead to reduction in visibility. This means the Hong-Ou-Mandel effect is a good indicator of distinguishability of single photons, but it provides a limited insight into their quality. Another shortcoming of the Hong-Ou-Mandel effect is related to the technological aspect of the employed single photon detectors. These particle detectors have a very broad detection spectrum incompatible with the much narrower spectrum of continuous homodyne detectors, which play crucial role in CV quantum information processing with light \cite{hybrid}. This discrepancy can be removed by using optical filtering, but this comes with the cost of detection efficiency and alteration of the photon statistics. Another issue is the fundamental identity of the single photon detectors as \textit{intensity} detectors. As such, the measurement is insensitive to the vacuum and can not reveal the wave portion of the interference - the relative phase between the modes -  related to the phase sensitive correlations between the field quadratures. As a consequence, with the particle detectors the state (\ref{state_id}) can not be differentiated from a completely mixed state $(|20\rangle\langle 20| + |02\rangle\langle 02|)/2$. Needless to say, in CV and hybrid quantum information processing this distinction is quite crucial.

We are therefore looking for a strong unifying witness capable of verifying the quality of single photon states as well as their indistinguishability. 
At the same time, we want a witness which employs high quality homodyne detectors in order to make it compatible with CV quantum information processing \cite{ourjoumtsev,Etesse}. And finally, we want to observe the interference effect directly, without the need to consider different measurement bases and settings, and subsequent reconstruction of the state's density matrix.

\begin{figure*}[th]
\includegraphics[width=400px]{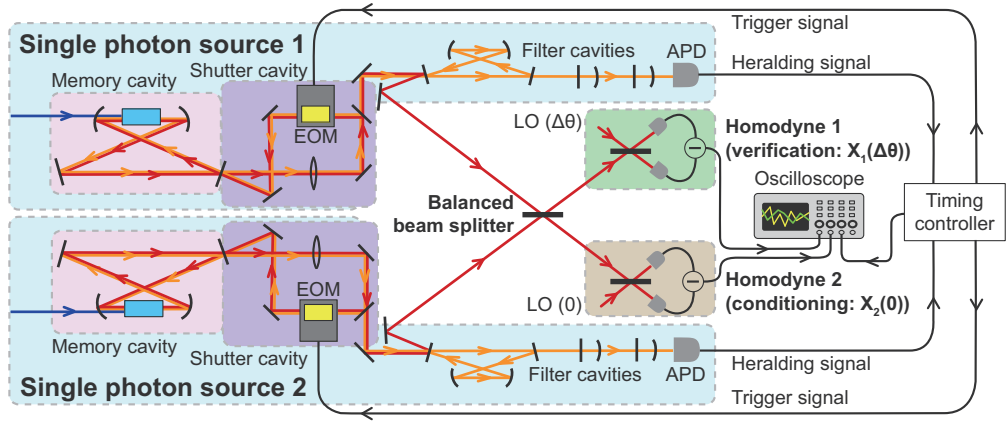}
\caption{{(Color online) Experimental setup. Two single photon states are independently generated and stored in the memory cavities. After synchronization, they are released at the same time, let to interfere on the balanced beam splitter, and recorded by the pair of homodyne detectors.   EOM: electro-optic modulator, LO: local oscillator, APD: avalanche photo diode.}}
\label{fig_ExpSetup}
\end{figure*}

All of these requirements are reconciled in a three step procedure, in which the two single photon states are first phase randomized, then they interfere on a balanced beam splitter, and finally
they are subjected to
the joint measurement of phase-correlated quadrature operators $X_1(\theta_1)$ and $X_2(\theta_2)$, where $X_j(\theta) = X_j \cos\theta + P_j\sin\theta $ with $j = 1,2$.
$X_j$ and $P_j$ obey canonical commutation relation $[X_j,P_j] = i$.
Because of the phase insensitivity of initial single photon states,
a common component of phases has no significance.
Hence, quadrature operators can be expressed only with the differential component $\Delta \theta = \theta_1 - \theta_2$ as $X_1(\Delta \theta)$ and $X_2(0)$.
Let us look at the ideal scenario with the state (\ref{state_id}) and consider a homodyne measurement of mode 2 yielding a value $x_2$.
This measurement projects the state of the remaining mode 1 into a state proportional to $e^{i\phi} \langle x_2 |2\rangle |0\rangle - \langle x_2|0\rangle |2\rangle$.
This state has zero mean values of the quadrature operators $X_1$ and $P_1$.
The remarkable thing is that by choosing some particular values of $x_2$, the state in mode 1 is projected into a squeezed state, whose second moment at a certain measurement angle shows a value less than $1/2$. The witness, confirming that state $\rho$ was obtained by interference of two phase insensitive single photon states on a balanced beam splitter, can be therefore formulated as
\begin{align}\label{directwitness}
    E[X_1^2(\Delta \theta)|X_2(0) = x_2] 
     = \frac{\mathrm{Tr}[ X_1^2(\Delta \theta)\otimes|x_2\rangle\langle x_2| \rho]}{\mathrm{Tr}[ 1\otimes|x_2\rangle\langle x_2| \rho]}    < \frac{1}{2},
\end{align}
where $E[\cdot|C]$ represents conditional mean value of an operator when condition $C$ is satisfied. The phase $\Delta \theta_\text{sq}$ that gives minimum second moment is tied with the phase of superposition $\phi$ as $\Delta \theta_\text{sq} = - \phi / 2$.
It is important that $\phi$ is arbitrarily controllable by classical phase locking, so $\Delta \theta_\text{sq}$ is known \textit{a priori} and can be set to an arbitrary value.
The procedure therefore requires only a single set of measurements, avoids any need for reconstruction, and (\ref{directwitness}) is a direct witness for both the indistinguishability and quality of single photons.
The conditioning on $x_2$ is related to the generalized theory of squeezing extraction presented in Ref.~\cite{zapletal}, but there are some interesting distinctions.
Due to the ideally vanishing nature of the first moments, the condition can be specified with help of the second \textit{noncentral} moment which makes it unsatisfiable by a lone single photon \cite{lonephoton}.
This is in contrast to observing squeezing of the conditional central moment, the variance, which can be achieved even for a single photon interacting with the vacuum and may therefore lead to false positives. However, the conditional squeezing can be considered as a witness for quality of \emph{single} photons \cite{zapletal}. For example, an imperfect single photon state represented by density operator $\rho = \eta|1\rangle\langle 1| + (1-\eta)|0\rangle\langle 0|$ can be confirmed as nonclassical for an arbitrary positive value of $\eta$. The method demonstrates phase-sensitive continuous variable nature of the correlations in photon anti-correlation experiments.
It is also interesting to point out that when the squeezing is replaced by another nonclassicality witness employing only single quadrature measurement, such as one proposed in Ref. \cite{vogel3}, no conditional nonclassicality can be detected.

The experimental setup for the test of the phase-sensitive Hong-Ou-Mandel effect for two single photon states is based on our former experiment reported in Ref.~\cite{MakinoHOM} and it is shown in Fig.~\ref{fig_ExpSetup}.
Each of the two photon sources is based on the heralded scheme with parametric down conversion contained in an optical cavity.
Photon pairs are generated from the continuous-wave pump light at 430~nm at random times.
One photon of the pair immediately leaves the cavity and is detected by an avalanche photodiode, thus heralding presence of another 860~nm photon stored in the cavity. A technique involving concatenated shutter cavity \cite{Yoshikawa, YoshikawaBook} can be then used to release the second photon at a desired time. This allows timing synchronization of the two generated photons and enables their interference on the balanced beam splitter.
With this technique, the simultaneous emission occurs 15 times in a second with 300 ns synchronization window and we gathered 12,000 events within an hour. Since the single photon states are generated independently,  their phase is random and needs no further randomization.
The state after interference is measured by phase-sensitive homodyne detection in both output modes of the balanced beam splitter: one mode is used for conditioning to generate squeezing and the other mode is used for the verification of squeezing in the conditioned state.
The relative phase between a local oscillator and the output mode at both homodyne detections, and the relative phase of two local oscillators which is equivalent to $\Delta \theta$ are actively stabilized.
The quantum efficiency of homodyne detectors and the visibility between a signal beam and the local oscillator are both more than 99\%.
Any correction of losses is not applied to the results below.

\begin{figure}[thbp]
\includegraphics[width=240px]{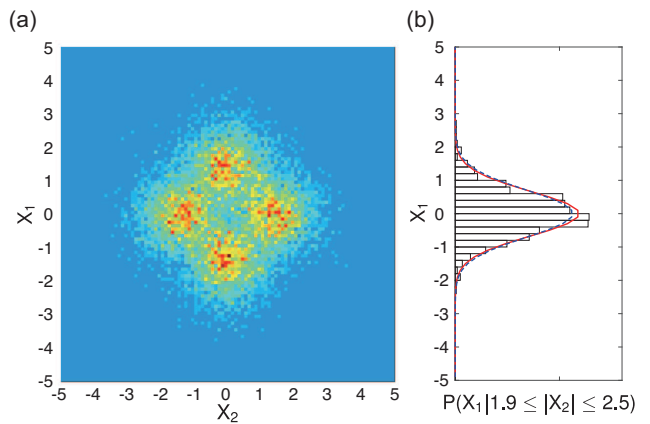}
\caption{{(Color online) Experimentally measured quadrature distribution of the state after interference with $\Delta \theta = \Delta \theta_\text{sq}$.
(a) Joint probability distribution $P(X_1,X_2)$.
(b) Normalized quadrature distribution $P(X_1)$.
An occurrence rate of $X_1$ with the window $1.9 \leq |X_2| \leq 2.5$ is shown in histogram.
Lines are Gaussian distributions with the variance of conditioned $P(X_1)$ (solid red) and that of vacuum fluctuation for reference (dashed blue), respectively.}}
\label{fig_hist}
\end{figure}
Fig.~\ref{fig_hist} (a) shows the joint measured distribution of quadratures $X_1$ and $X_2$ of the two modes emerging from the beam splitter with $\Delta \theta = \Delta \theta_\text{sq}$.
The quadratures were locked against each other, but their relation to the outside reference varied over time. This effectively resulted in a statistics which is equivalent to measurement without an outside phase reference. We will therefore omit the phase dependence from now on. We can see that the quadrature data are distinctively correlated in a manner consistent with the target state. In accordance with the proposed witness, we are looking for a quantitative indicator of the Hong-Ou-Mandel coherence in the form of the second conditional moment  $E[ X_1^2|X_2 = x_2]$. The theory asks for conditioning on a single measurement result. Since this is impossible, we have relaxed the requirement for condition and accepted situations in which the value of quadrature $X_2$ fell into a bin of width $\Delta$, $x_2-\Delta/2 \le X_2 \le x_2+\Delta/2$. Due to the symmetrical nature of the distribution, we also used both positive and negative values in the post-selection.
An example of the conditional quadrature distribution of $X_1$ is shown in Fig.~\ref{fig_hist} (b). The conditioning window, denoted by set $B$, was set to $1.9 \leq |X_2| \leq 2.5$ and it picked 1,028 of the initial 12,000 data points. The conditional variance was found to be $E[ X_1^2|X_2 \in B ]  = 0.45 $, which is smaller than the vacuum fluctuations variance of 0.5.

\begin{figure}[thbp]
\includegraphics[width=240px]{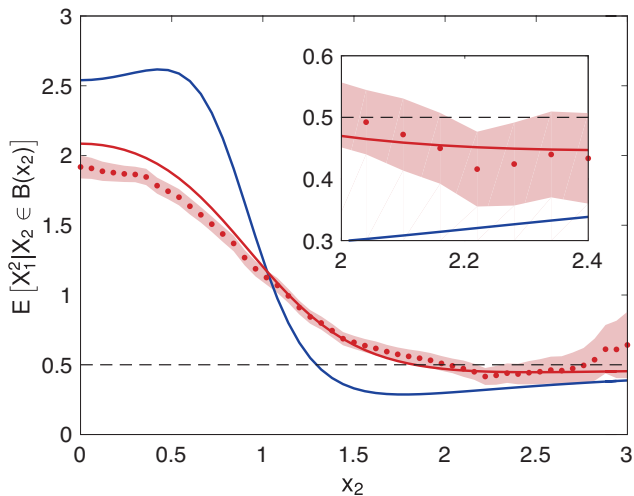}
\caption{(Color online) The conditional second moment $E[X_1^2| X_2 \in B(x_2)]$ where 
$B(x_2)$ is the set of values satisfying $x_2-\Delta/2 < |X_2| < x_2 + \Delta/2$.
The width of the post-selection interval was chosen to be $\Delta = 0.6$. Dashed blue line marks theoretical prediction for the interference of two ideal single photon states. Solid red line shows prediction for interference of two imperfect photons produced by the cavities. Red dots represent experimental data with the light red zone marking the $3\sigma$ confidence band. The inset shows detailed behavior in the area with the highest certifiable squeezing.   }
\label{fig_condVX2}
\end{figure}

Let us now discuss the results in more detail, giving more attention to possible statistical errors, and make sure the observed squeezing is significant. For that we need to construct the confidence band for the estimated variance. Under the assumption that the conditional distributions are approximately Gaussian we used Monte Carlo method to simulate 1,000 runs of the experiment, estimated their moments and evaluated their statistics. The standard deviation of the estimated variances depends on the number of data points and therefore on the position and the width of the post-selection interval. We have optimized over these parameters and found minimal certified variance for conditional interval width $\Delta = 0.6$ (see Appendix A for details). The conditional variances for this scenario are shown in Fig.~\ref{fig_condVX2}. We can see that the measured data closely follow the curve predicted from knowledge of the two independent single photon states, which was obtained by their individual tomographic reconstructions \cite{MakinoHOM}. This shows that the amount of squeezing is limited mainly by the quality of the single photons and not their coherence. Furthermore, the squeezing is observable even if we take the three standard deviations wide confidence band into account.  This confirms that the squeezing is not a statical artifact and demonstrates both the nonclassical nature of the two initial single photon states and their mutual coherence. This first application of the proposed witness also shows its sensitivity to small imperfection of these components. Note, visibility of the traditional Hong-Ou-Mandel interference estimated from the tomography data would be $V=0.7$ in our experiment.

In summary, we have suggested a phase sensitive Hong-Ou-Mandel effect as a witness for recognizing high quality of individual photons and their ability to quantum mechanically interfere. Such interference is the basic building block of future hybrid optical technologies which combine both continuous and discreet aspects of light \cite{hybrid}. The suggested witness tests both of these properties and it does that directly, using only a single set of measurements without needing a sequence of individually incompatible measurements and without directly exploiting entanglement \cite{entanglement} or discord \cite{discord}. We have applied the witness to the pair of on demand generated photons and demonstrated both their nonclassical nature and their mutual coherence. The witness can be directly applied to other experimental platforms where generation of interfering Fock states plays an important role, like atomic ensembles \cite{atomic}, optomechanics \cite{optomechanics1,optomechanics2}, and superconducting circuits \cite{superconducting1, superconducting2}.

\section*{Appendix A}
\begin{figure}[th]
\includegraphics[width=250px]{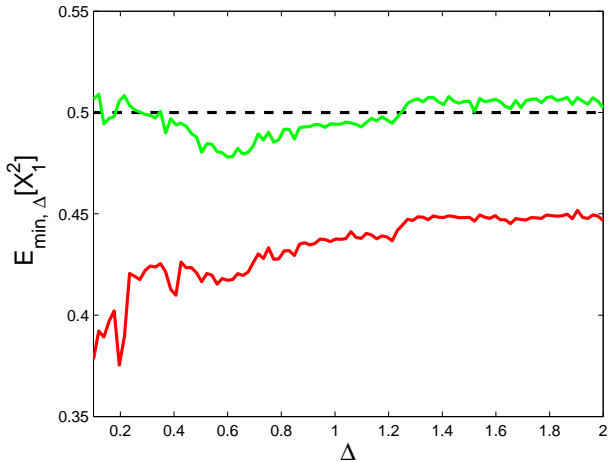}
\caption{(Color online) The conditioned second moment $ E_{\mathrm{min,\Delta}}[X_1^2]$ (red) and its upper confidence band (green) relative to the width of the postselection interval $\Delta$. The position of the conditioning interval given by the measured value of $X_2$ was optimized over in order to obtain minimal certifiable variance given by the upper confidence interval.  }
\label{fig_Delta}
\end{figure}

The conditional second moment of the $X_1$ quadrature depends on the width $\Delta$ and the position of the conditioning interval. A narrow interval generally yields the lowest possible moment values, but they come at the cost of low number of data points and, consequently, lesser certainty about the validity of the result. The best certifiable violation is obtained when the upper boundary of the confidence band reaches its minimum value. Fig.~\ref{fig_Delta} shows the second moment with its upper confidence band estimated from the measured data. For all considered values of $\Delta$ we have optimized over the values $x_2$ marking the center of the interval to obtain the minimum of the upper confidence band boundary.
The plotted moment can be therefore expressed as
\begin{equation}\label{}
    E_{\mathrm{min,\Delta}}[X_1^2] = \min_{x_2} \left\{ E\Bigl[X_1^2\Big| x_2 - \frac{\Delta}{2}< |X_2| < x_2 + \frac{\Delta}{2}\Bigr] \right\}.
\end{equation}
We have obtained the lowest certifiable variance for $\Delta = 0.6$.

\section*{Acknowledgement}
This work was partly supported by CREST of JST, JSPS KAKENHI, APSA.
P. M., P.Z., and R. F. acknowledge Project GB14-36681G of the Czech Science Foundation.
Y. H. acknowledges support from ALPS.
K. M. acknowledges support from JSPS.


\begin{thebibliography}{cc}
\bibitem{lonephoton}
Single photon divided on a beam splitter and conditioned by homodyne detection can be always expressed as $c_0|0\rangle + c_1|1\rangle$. In this state, $\langle X^2\rangle = 0.5 + |c_1|^2$ and the moment is never squeezed.
\end{thebibliography}
\end{document}